\pdfoutput=1

\documentclass[manuscript]{aastex}

\newcommand{\beq}{\begin{equation}}
\newcommand{\eeq}{\end{equation}}

\newcommand{\kms}{\mbox{ km s$^{-1}$}~}

\newcommand{\Mo}{\mbox{M$_{\odot}$}~}

\newcommand{\Lo}{\mbox{$L_{\odot}$}~}
\newcommand{\Ro}{\mbox{R$_{\odot}$}}

\shorttitle{Rotating Stars and Bipolar Nebulae}
\shortauthors{Garc\'{\i}a-Segura et al.}

\begin{document}

\title{Rotating Stars and the Formation of \\  Bipolar Planetary Nebulae II:  Tidal Spin-up}

\author{G. Garc\'{\i}a-Segura}
\affil{Instituto de Astronom\'{\i}a, Universidad Nacional Aut\'onoma
de Mexico, Km. 103 Carr. Tijuana-Ensenada, 22860, Ensenada, B. C., Mexico}
\email{ggs@astrosen.unam.mx}

\author{E. Villaver}
\affil{Departamento de F\'{\i}sica Te\'orica, Universidad Aut\'onoma de Madrid,
       Cantoblanco, E-28049 Madrid, Spain}

\author{A. Manchado\altaffilmark{1,2}}
\affil{Instituto de Astrof\'{\i}sica de Canarias,  Via L\'actea s/n, E-38200 La Laguna, Tenerife, Spain}

\altaffiltext{1}{Departamento de Astrof\'{\i}sica, Universidad de La Laguna,
E-38206 La Laguna, Tenerife, Spain}
\altaffiltext{2}{Consejo Superior de Investigaciones Cient\'{\i}ficas (CSIC), Spain}

\author{N. Langer }
\affil{Argelander-Institut f\"ur Astronomie, Universit\"at Bonn, D-53121 Bonn, Germany}

\author{S.-C. Yoon}
\affil{Astronomy Program, Department of Physics and Astronomy,
Seoul National University, Seoul, 151-747, Republic of Korea}

\begin{abstract}
We present new binary stellar evolution models that include the effects of tidal forces, 
rotation, and magnetic torques with the goal of testing Planetary Nebulae (PNe) 
shaping via binary interaction. 
We explore whether tidal interaction with a companion 
can spin up the AGB envelope. To do so we have selected  binary systems with 
main sequence masses of 2.5 \Mo and of 0.8 \Mo and evolve them allowing 
initial separations of 5, 6, 7, and 8 AU.  The binary stellar evolution models 
have been computed all the way to the PNe formation phase  or 
until Roche lobe overflow (RLOF) is reached, whatever happens first. 
We show that with initial separations of 7 and 8 AU, the binary avoids
entering into RLOF, and the AGB star reaches moderate rotational velocities 
at the surface ($\sim 3.5 $ and $\sim 2 $ \kms respectively) during the 
inter-pulse phases, { but after the thermal pulses it drops to a final
rotational velocity of only $\sim 0.03 $ \kms}.
For the closest binary separations explored, 5 and 6 AU, 
the AGB star reaches  rotational velocities of $\sim 6 $ and $\sim 4 $ 
\kms respectively  when the RLOF is initiated.
We conclude that the detached { binary} models that avoid entering the RLOF phase during 
the AGB will not shape bipolar PNe, since the acquired angular momentum is lost 
via the wind during the last two thermal pulses.  
{ This study rules out tidal spin-up in non-contact binaries as a 
sufficient condition to form bipolar PNe}.

\end{abstract}



\section{INTRODUCTION}

During nearly 30 years of research in the formation of planetary
nebulae (PNe), several explanations have been proposed in the literature to produce the
bipolar shapes. The most popular one involves the presence of a dense medium in the equatorial
latitude, which collimates the later fast stellar wind
(Calvet \& Peimbert 1983; Balick 1987; Icke 1988; Icke et al. 1989; Mellema 1991;
 Frank \& Mellema 1994).
A number of models, have the focus on how to form a dense medium along the equator
of the asymptotic giant brach (AGB) star by stellar rotation
(Ignace et al. 1996; Garc\'{\i}a-Segura 1997; Garc\'{\i}a-Segura et al.1999),
while others concentrate on magnetic forces,
succeeding only sometimes at creating such a dense medium (Matt et al. 2000).

A long debate regarding the need of a binary companion has also been taking place during
these last three decades, since AGB wind asphericities could naturally arise from the interaction
of AGB stars with a companion in several ways.
The most popular ones involve common envelope
evolution and the initial phase of  spiral-in  (Livio 1993; Soker 1997; Sandquist et al. 1998;
Nordhaus \& Blackman 2006; de Marco 2009; Ricker \& Taam 2012; de Marco et al. 2013),
gravitational focusing (Gawryszcak et al. 2002),  wind-capture disks 
(Huarte-Espinosa et al. 2013) { and Wind-Roche-Lobe-Overflows 
(Podsiadlowski \& Mohamed 2007; Mohamed \& Podsiadlowski 2012)} .
Along these lines, in our paper I
(Garc\'{\i}a-Segura et al. 2014) we provided a solid probe that
bipolar PNe formation cannot result from the evolution of a single AGB star, given 
that these stars do not carry the necessary  angular momentum, or to be more specific, 
the necessary surface rotation.

In this second paper, we explore the alternative of binary (versus single) evolution for
bipolar PNe formation from the view point of stellar evolution models.
We calculate the effects that a secondary star
has onto the primary stellar surface and see if they can explain the
formation of a bipolar nebula.  To do so, we compute the first binary models of AGB stars allowed to be
spun-up by tidal interactions,  and study  which rotational velocity
we can get using this approach. As we mentioned in paper I, surface rotational velocity
values above 2 \kms could in principle produce considerable asymmetries in these stars.

The spin-up of stellar envelopes has only been calculated for massive stars
(see Detmers et al. 2008) to explain the formation of gamma-ray bursts.
We recall however that although the present calculations are the first in the
context of the formation of bipolar PNe, the idea has been around for quite some time.
Livio (1994) first mentioned tidal spin-up as a possible origin of bipolar PNe and later on Soker (1998)
suggested  that as a consequence of the spin-up, the primary's envelope could rotate at several percent
of the breakup velocity, and this will probably result in a small, but non-negligible,
effect on the mass-loss geometry. 
In this paper we precisely focus on above ideas, and compute how important spin-up is in this 
context. The outline of the paper is as follows: \S~2 describes the numerical scheme and 
physical approximations as well as the inputs in our calculations; \S~3  shows the results 
of the stellar evolution calculations; and finally in \S~4 we disscus the results and provide 
the main conclusions of this study.

\section{STELLAR MODELS, METHODS AND PHYSICAL ASSUMPTIONS}

The stellar evolution calculations have been done using the Binary Evolution
Code (BEC) (Petrovic et al. 2005; Yoon et al. 2006).  BEC is a one-dimensional
hydrodynamic, stellar evolution code designed to evolve stellar models of
single and binary stars, which { originates from the STERN code (Langer 1991). 
The structure and evolution of stars is governed by a set of partial
differential equations, the so-called stellar structure equations. In BEC,
which uses the hydrodynamic version, the inertia terms are used (see Kozyreva et al. 2014).
This code includes diffusive mixing due to convection, semi-convection (Langer
et al. 1985), and thermohaline mixing as in Wellstein et al. (2001) and
Cantiello \& Langer (2010) . } 

{ In rotating stars, centrifugal forces act on the stellar gas leading
to deviations from spherical symmetry. For slow to moderate rotation these
deformations remain rotationally symmetric (Tassoul 1978), and no
triaxial deformations are expected. The shapes of equipotentail surfaces
are affected by the centrifugal potential and therefore deviate from 
spherical symmetry. In this scenario, the momentum equation and the energy
transport equation for spherical symmetric stars have to be modified.
To take this effect into account,  
the effect of the centrifugal force on the stellar structure is calculated
following the method of Kippenhahn \& Thomas (1970) in the approximation
of   Endal \& Sofia (1976), and applied to the hydrodynamic stellar structure
equations (Heger et al. 2000). 
Thus, instead of spherical shells, the mass shells
correspond to surfaces of constant pressure.
The corrections are applied to the
acceleration and the radiative temperature gradient.
According to Zahn (1975), Chaboyer \& Zahn (1992) and Zhan (1992), 
anisotropic turbulence acts much stronger on isobars than in perpendicular directions.
This enforces shellular rotation rather than cylindrical rotation, and it
removes compositional differences on isobaric surfaces. Therefore, it
can be assumed that the matter on such surfaces is chemically
homogeneous. Together with the shellular rotation, this allows to retain
the one-dimensional approximation (Heger et al. 2000). }  

Time-dependent chemical mixing and transport of angular momentum due to 
rotationally-induced instabilities, including the shear instability and the 
Golreich-Schubert-Fricke instability,
and the Eddington-Sweet circulations are also included as a diffusive process
(Heger et al. 2000).  We also include the transport of angular momentum due to
magnetic fields (Spruit 2002), as in Heger et al. (2005) and Petrovic et al.
(2005).  
{ They considered dynamo action in radiative layers, resulting from winding up 
magnetic fields due to differential rotation, that creates toroidal fields, 
and generation of poloidal fields from toroidal fields due to the Tayler 
instability (the Talyer-Spruit dynamo). 
The strength of magnetic fields  is calculated using the steady state solution 
of the Tayler-Spruit dynamo, which is given as a function of the degree of 
differential rotation.   The magnetic field therefore is instantaneous and
has no memory of previous timesteps.  
In our cases, the strongest magnetic field is found in the 
boundary layer between the core and the envelope where the degree
of differential rotation is the largest, while in the rest of the envelope, 
magnetic fields are very weak.}
The rotation profile in the convection zone is self-consistently
calculated by solving the diffusion equation for the transport of angular
momentum.

{ The physics of binary interactions was implemented by Braun (1997, PhD Thesis) 
and Wellstein (2001, PhD Thesis).  We assume a circular orbit and adopt the Eggelton's
approximation for the Roche-lobe radius (Eggelton 1983), for our considered
binary systems. We calculate the mass transfer rate due to Roche-lobe overflow
following the method by Ritter (1988), in an implicit way 
(Note, however, that our models do not involve mass transfer in this article). 
We adopt the prescription by Podsiadlowski et al. (1992) to calculate the evolution of 
the binary orbit resulting from mass transfer and stellar winds mass loss, and the
numerical result by Brooksaw and Tavani (1993) to calculate the amount of
specific angular momentum of the stellar wind material that escapes the binary
system. }

The synchronisation of the binary system is computed with the same method as 
in Wellstein (2001) (see also Detmers et al. 2008), using the synchronisation 
time given by Zahn (1977) for convective stars :
\beq \tau_{\rm sync}  \,\, \propto  \,\, q^{-2} \,\, \left( {d \over R} \right)^6 \,\,, \eeq
where  $q \,\, = \,\, M_{\rm 2}/M_{1}$ is the mass ratio of the binary, with $M_{1}$
the mass of the primary star that evolves up to the AGB phase, 
and $M_{\rm 2}$ the mass of the secondary main sequence star,
 $d$ is the orbital separation, and $R$ the radius of the primary AGB star.
The angular momentum exchange $\Delta J$ that is added to the { whole} AGB star at 
each time step $\Delta t$ due to tides is
\beq \Delta J \,\, = \left( J_{\rm AGB} - J_{\rm sync} \right) \left( 1 - {\rm e}^{-\Delta t/\tau_{\rm sync}} \right) \,\,, \eeq
where $J_{\rm sync}$ is the spin angular momentum that the AGB star
would have when the system is in synchronous rotation.

{ Recently, Paxton et al. (2015) implemented binary physics into the  MESA code,
by including many ingredients from BEC. Since the two codes use very similar 
physics, a comparison of the two would not be too meaningful 
(see Section 2.6 for Numerical Tests in Paxton et al. 2015). 
There exists no other code that includes both binarity 
and differential rotation at the present.  
The accuracy of our results  is more dependant on the 
synchronization physics rather than on numerics, because the numerical implementation 
is quite straightforward.  
Along these lines, there two articles which are useful to mention. In the first one,
Khaliullin \& Khaliullina (2010) found that Zahn's theory is consistent with 
observations of 101 eclipsing binaries with early-type main-sequence component,  
and in the second one, Zamanov et al. (2007) reports evidence 
for synchronisation of 29 S-type symbiotic stars, which are similar type of binaries. 
The assumption of circular orbits is a good approximation in our case, since
the  circularization timescale is comparable to synchronization timescale.  
In the case that the orbits were very eccentric, there might be episodes of  mass transfer
when the stellar components become close enough, for a given mean orbital separation. 
Considering such a situation would be very interesting, but will be the subject of  
future work (see Staff et al. 2015 for a recent work along these lines).}

The majority of Solar-type stars did seem to have stellar companions 
(Abt \& Levy 1976; Duquennoy \& Mayor 1991).
The fraction of main sequence binaries was found to be in the range from 65 \% to 100 \%
with only about 30\% G-dwarf primaries having no companion up to about 0.01 M$_{\odot}$ 
(Duquennoy \& Mayor 1991).
The multiplicity fraction however seems to have a dependency on the mass of the primary with 
binary systems being about 50 \%
between spectral types B4 and A7 (Kouwenhoven et al. 2007) and about 33 \% for F6-K3 stars 
using the more recent estimates by Raghavan et al. (2010). 
{ These values are consistent
with the fraction of binary systems found in central stars of PNe (Douchin et al. 2015).
Note that both the binary period distribution and the binary fraction are 
meaningful in doing this comparison}. 
 
In the current study, we use binary systems where the primary star have an initial
mass of 2.5 M$_{\odot}$.  The main reason that motivate
the choice of the initial progenitor mass is the higher mass stellar progenitor that the 
bipolar PNe morphological class seem to have, as indicated by their chemical abundances 
(see e.g. Stanghellini et al. 2006; Manchado 2004),
and their closer distribution to the galactic plane (Calvet \& Peimbert 1983; 
Corradi  \& Schwarz 1995,
Manchado 2004; see also paper I Garc\'{\i}a-Segura et al. 2014).
For the binary system, we have chosen as the secondary a stellar companion with
0.8 M$_{\odot}$. These stars represent  one of the most numerous stars on the main sequence and have been
found to be numerous as well as secondaries in the study of  Sco OB2  by Kouwenhoven et al. (2005).
A mass fraction $q \,\, = \,\, M_{\rm 2}/M_{1}$ of $0.32$ seems to be reasonably likely and consistent
with the observed values by Kouwenhoven et al. (2005). Note that Duquennoy \& Mayor (1991) also found 
that the binary mass distribution peaks around $q= 0.3$ for Solar-type stars concluding that in 
general binaries can be formed by association of stars build from the same IMF. The secondary star 
is treated as a point mass in this study.

For the stellar mass-loss rate we adopted Reimers (1975) with $\eta =0.5$ during the Red Giant phase and the
Vassiliadis \&  Wood (1993) parameterization for the AGB phase.

We have computed two sets of models (8 in total) from the zero age main sequence (ZAMS)
 to the AGB phase using primaries with two different initial stellar rotation rates
and  four different binary initial orbital separations. The binary parameters of all the computed models at the ZAMS are listed in Table 1.
In the first set of models we consider a very small initial rotation, nearly zero, which is the rotation
that would result from synchronization at the zero age main sequence (ZAMS).
For the second set of models we have adopted an initial equatorial rotation velocity
of 250 \kms at the ZAMS  representative of single stars (Fukuda 1982) with the mass of the primary.
This velocity has been chosen as well to allow comparisons with the results
from paper I. The initial orbital separations used in the calculations are 5, 6, 7 and 8 AU.
To make the choice for the separations of the binaries, 
we computed first preliminary basic models that allowed us to select two cases with RLOF 
and two cases that avoid RLOF,  as it will be discussed in the next section. 
The minimum binary separation of 5 AU has been tuned to ensure that binary interaction 
is achieved during the late AGB phase, avoiding altogether a strong tidal interaction
during both the main sequence and the Red Giant Branch phase.

\section{RESULTS}

The evolution of the surface equatorial rotational velocity during the thermal pulsing AGB phase 
is shown in Figure 1, for each of the initial binary parameters listed in Table 1. We do not show 
the earlier evolution because the tidal spin-up is not important as  the synchronization time is 
quite dependent of the radius (see equation 1).  Only when the stellar radius during the AGB
is large enough, for a given binary separation, tidal forces become important and the 
synchronization time is short enough that the spin-up becomes evident.
Figure 1 also shows for comparison (dotted line) the results of the model presented 
in paper I in which we included angular momentum transport induced by magnetic torques 
inside the star. 

We see in Figure 1 that all the binary models
have larger surface rotational velocities than the single stellar models computed in paper I 
(where surface rotational velocities larger than  $1$  \kms were never reached at the late
thermal pulsing AGB phase).
We find that the models with initial separations of 5 and 6 AU end up in RLOF phases while
the models with initial separations of 7 and 8 AU avoid RLOF altogether.
The  AGB stars, in the more separated binary models, reach moderate rotational 
velocities at the surface ($\sim 3.5 $ and $\sim 2 $ \kms for models with 7 and 8 AU 
respectively) during the inter-pulse phase, just before the last two thermal pulses, and 
for the closest binary separations explored, 5 and 6 AU, 
the AGB stars reach rotational velocities of $\sim 6 $ and $\sim 4 $ \kms respectively,  when 
the RLOF is initiated. Note that the models that do not
enter { in} the RLOF phase lose most  of the gained angular momentum via the wind during the last 
two thermal pulses. This is clearly reflected in Figure 1 with the quick drop of velocity 
at the end of their AGB evolution.

We find that the maximum rotation velocity gained at the
thermal pulsing AGB  phase does not depend of the initial velocity assumed for the primary, 
but only on the initial separation of the binary. To illustrate this  we plot the spin angular momentum 
of the primary star in Figure 2.  We find that the final spin angular  
momentum reached by the star is independent of the initial assumption, since the major 
contribution to the spin angular momentum of the primary comes from the orbital
angular momentum. Although there is an important transfer of orbital angular momentum to the
primary spin angular momentum, the separation (and the orbital period) growth in time
due to mass-loss, as it is seen in Figure 3. This figure also shows that
models c5 and r5 have almost reach corotation at the time our computation is stopped
(because they reach RLOF). Models c6 and r6 are slightly below the corotation speed.

One of the most important results of the comparison allowed by our models is that the 
binaries that do not reach RLOF lose the gained angular momentum by mass-loss at the 
end of the AGB phase. This means that for these models the formation of
bipolar nebulae via magnetohydrodynamical collimation is not possible, since at the time when
a PN is formed, after an uncertain amount of transition time, the stellar surface rotation 
is very small,  and the effect that the secondary might have had in speeding up the primary 
will be negligible due to the large separation.  Since the sound speed
at the surface of an AGB star is of the order of $\approx$ 1 km s$^{-1}$, 
values below that for the rotation will not have any impact on the wind compression mechanism
(Bjorkman \& Cassinelli 1993; Ignace et al. 1996). When the calculations are stopped,  
we obtain surface rotational speeds
of 0.03, 0.03, 0.4 and 0.09 \kms  for models c7, c8 , r7 and  r8 respectively. The 
stellar parameters are shown in Table 2.

On the other hand, the models that reach RLOF generate, for a timespan that requires further calculations, 
the necessary dynamical conditions to form an equatorial density enhancement in the circumbinary medium, 
since the rotation is fast ($\sim$ 6 \kms)  and most likely sustained for the necessary time.  
These models are promising in forming bipolar nebulae as we will discuss in the next section.
Table 3 gives the parameters  of the binary systems approaching the RLOF phases at the time the 
computations are stopped.

\section{DISCUSSION AND CONCLUSIONS}

Our set of binary calculations shows that it seems a very reasonable assumption 
that the models that avoid RLOF phases will not form bipolar PNe, since all the 
gained angular momentum during the AGB phase will be lost via the wind in the last 
two thermal pulses.  It is important to note as well the fact that the whole timescale 
affected by this process is very long, of the order of 100,000 \, yr,
taking into account the last thermal pulse, and the possibility of a few thousand 
years { more } of uncertain transition time until the central star is hot enough for the fast 
wind mechanism to operate and shape the nebula (see e.g. Villaver et al. 2002).  
Any fingerprint of a transient, equatorial density enhancement will be dissipated at 
the time of PN formation in these models.  Then, only the models that reach the RLOF { phases} 
are promising to explore bipolar PNe shaping, since their evolution will proceed 
differently than those for single stars or non-contact binaries.

Following our results, it will be extremely useful to have the distribution
and latitudinal dependence of the stellar wind coming out from the binary
system, in order to model the formation of a bipolar PNe. Unfortunately,
the equations by Bjorkman \& Cassinelli (1993) and Ignace et al.(1996) that study the
formation of wind-compression zones and wind-compression disks around single stars
cannot be applied directly in this case, since the geometry is different for
binary systems and the center of mass of the binary is not at the center of the AGB star.  
Currently, there is not any analytical or semi-analytical study that predict  
the distribution of gas in the nearby circumbinary  medium for these binary stars, 
except for the studies where the outflow comes from the Lagrange point ${\rm L}_2$ 
(Shu et al. 1979), as we will discuss below.

We have made a rough estimate on how important a wind compression mechanism could be 
using the results of our models and keeping in mind the applicability restrictions 
outlined above.  We focus the discussion in the last output from model c5. In this 
model the AGB star is at  19 \% of its critical rotation speed, which according to 
Ignace et al.(1996) can be translated into a density contrast ratio equator/pole  of $\sim$ 9.
Note that these numbers are given in the non-inertial reference frame of the AGB star but we can 
translate them to the inertial, reference frame at the center of mass.
The total mass of the binary system is $3.006$ \Mo (accounting for the mass lost) 
and the center of mass is located at $306$ \Ro \,  from the center of the AGB star, 
i.e., $\sim$ 26 \% inside of its envelope. With this is mind, the equatorial surface at 
the  opposite side of the secondary has a velocity of 14 \kms, which translates into 
close to $\sim$ 50 \% of the critical rotation at that location. A  50 \% of the 
critical rotation is within the range of conditions for the wind-compression disk 
formation, of course if the equations of Ignace et al.(1996) could be applied in the 
scenario we have just described.  What this exercise emphasizes is that mass-loss is 
expected to be important at the opposite side from the secondary, in the direction of 
the Lagrange point ${\rm L}_3$, where the escape velocity at the surface  is smaller 
than the value that an isolated star would have. This mass-loss enhancement is expected to 
occur with the shape of a spiral arm.

It is also important to mention that the system will lose an important amount of mass
through the Lagrange point ${\rm L}_2$.
This mass-loss is expected to have the shape of a spiral arm as well, since the mass ratio 
${\rm q}  = 0.36 $ at the beginning of the RLOF phase is  inside the range 
$ 0.064 \leq  q \leq  0.78 $  (see e.g. Shu et al. 1979; Mohamed \& Podsiadlowski 2012; 
Pejcha et al. 2015).  It is also interesting to note in this context that the ratio 
 ${\rm q}$ increases along the subsequent evolution as the primary looses mass and the 
secondary accretes part of it.  Once that the ratio  ${\rm q}$ exceeds the value of 0.78, 
and if the system is still in RLOF, the ejection of gas { through ${\rm L}_2$ } 
will piles up at a finite radius and will form a bounded mass-loss ring instead of an 
unbound outflow (Shu et al. 1979). 

How much time the system will remain in the RLOF phase depends on the rate at which 
{ the AGB radius expands, how soon the common envelope is formed}, the rate at which
orbital angular momentum is lost (shrinking the orbit of the binary system) and on 
the mass-loss rate from the binary system (increasing the orbital distance).  
In any case, the proper computations of the mass-loss through the Langrangians 
points ${\rm L}_2$ and ${\rm L}_3$ is challenging.  These new scenarios are of great 
interest for future hydrodynamical computations for the formation of bipolar PNe such as the 
ones recently carried out by Staff et al. (2015).

From our binary stellar evolutionary models  with an initial ${\rm q}  = 0.32 $, 
we  find that the maximum rotation velocity  during
the AGB phase, as a result of tidal spin-up, is independent of the initial rotational 
velocity at the ZAMS, and only depends on the initial  separation.

We find very unlikely that binaries which avoid the RLOF phases 
could form bipolar PNe with an equatorial waist. The angular momentum lost by the stellar 
winds in the last thermal pulses, and the unknown transition time from the AGB to the 
PN formation will dissipate any transient, equatorial density enhancement formed in the nearby 
circumbinary medium.
The same reasoning applies to the engulfment of giant planets if they do not
produce the ejection of the envelope, since no matter
how much angular momentum is gained by the AGB star, the last thermal pulses will carry it
away via the stellar winds.

{ The above discussion is focussed only in the classical formation of bipolar nebulae,
in which a slow, equatorially denser wind inhibits the future expansion of a fast wind 
in the equatorial regions.  However, this is not the only model that accounts for the 
formation of bipolarity, since jets and collimated outflows  may form around the companion 
(Soker \&  Rappaport 2000; Huarte-Espinosa et al. 2013) and power PNe, as they are known 
to do in Symbiotics (Corradi \& Schwarz 1993).
These scenarios are based on the formation of an accretion disk around the secondary star, 
where the ejected collimated gas forms the bipolar nebula,  before the formation of a 
fast wind by the central star.
The separation of the binary system is crucial here in determining the energy of the outflow.
According to the separation of the binary, the systems can be classified into three types:  
Wind-Capture disks for large separations;  Wind-RLOF disks for small separations,
but still larger than the distances required for RLOF; and the RLOF disks formed at the 
RLOF phases.  
Note that systems with small separations (such as models c5, c6, r5 and r6)  
will pass through the three different types consecutively, as the AGB radius increases 
during stellar evolution.
In the Wind-Capture disk, a companion orbiting through the dense wind of an AGB star will 
capture enough wind material to form an accretion disk and (most likely) power a jet
(Huarte-Espinosa et al. 2013).
The Wind-RLOF disk is a relatively new mechanism which allows far larger
orbital separations than traditional RLOF. High accretion rates are possible
with Wind-RLOF scenarios and the determination of its presence requires solutions for the AGB
wind structure (Podsiadlowski \& Mohamed 2007; Mohamed \& Podsiadlowski 2012).
Finally, we have the outflows from RLOF phases, where the amount of accretion is the largest one. 
Although neither of these possibilities are accounted for in the present study, they are not
exclusive, in the sense that the three types, specially the last two (Wind-RLOF and RLOF) 
will form also a density enhancement in the orbital plane with a spiral shape, due to 
the mass-loss through the Lagrange point ${\rm L}_2$.
In other words, the accretion disks will form both the jets and the density
enhancement in the orbital plane.

Since the orbital separation increases very fast at the last thermal pulses due to the
large mass-loss (Figure 3), the wind accretion is expected to decay with time in the first
two scenarios. Thus, it is very likely that the formation of bipolar PNe 
follows the evolution of binary systems with 
separations where RLOF phases are attained, as well as the subsequent (or imminent)
common envelope phases. Proper computations of the time spent in each of the three 
phases will be extremely important and will be  the subject of future studies.

Finally, another point which needs consideration in this study is the role of magnetic fields.
The creation of fields by companions in common envelope scenarios has been discussed 
in the literature and offers another route whereby binary stars produce bipolar PNe 
(Tout \& Reg\"os 2003;  Nordhaus \& Blackman 2006).  
Note that the spin-up of the envelope in the above studies is the direct product of a
spiral-in process during the common envelope evolution.  In our cases, however, 
the spin-up of the envelope is due to tidal forces.  In the case of
non-contact binaries, we found that magnetic fields are very 
weak inside of the envelope where the degree of differential rotation is negligible. 
The rotation velocity become too small at the end of the AGB phase, and it would not make 
any big difference from single stars.
We find very unlikely that magnetic fields could be of any importance here, 
which probably will have only a role in forming cool spots (Frank 1995; Soker 2001).

On the other hand, for the closest systems, it would make a big difference. 
Since AGB stars rotate  at $\sim$ 20--50 \% of their critical rotation,
this is significant enough to make an alpha-omega dynamo efficient. 
But, as we discussed above,  they will become a common envelope soon once mass transfer starts, 
and the later evolution would be governed by common envelope. 
So, probably, even in this case, the role of convective dynamo might be minor, although
this requires further investigation, since the accumulate magnetic energy could be
important for the ejection of the envelope.

In conclusion, this study rules out that tidal spin-up in non-contact binaries
could be a sufficient condition to form bipolar PNe.  }

\acknowledgments

G.G.-S. is partially supported by CONACyT grant 178253.
E.V. and A.M. work was supported by the Spanish
Ministry of Economy and Competitiveness, Plan Nacional de Astronom\'{\i}a y Astrof\'{\i}sica, under
grants AYA2014-55840-P and  AYA-2011-27754 respectively.
{ We would like to thank our anonymous referee for his valuable comments which improved the
presentation of the paper.}

\clearpage

\begin{table}
\begin{center}
\caption{Initial Parameters of Binary Stellar Models at ZAMS }
\begin{tabular}{ccccccc}
\tableline\tableline
Model &  d (A.U.) & $M_{1}$ (\Mo) & $M_{2}$ (\Mo) & $v_{\rm rot}$ (\kms) & Description  \\
\tableline
c5 & 5 & 2.5   & 0.8 &  $\sim 0$ & corotation  \\
c6 & 6 & 2.5   & 0.8 &  $\sim 0$ & corotation  \\
c7 & 7 & 2.5   & 0.8 &  $\sim 0$ & corotation  \\
c8 & 8 & 2.5   & 0.8 &  $\sim 0$ & corotation  \\
r5 & 5 & 2.5   & 0.8 &  250  & fast rotator   \\
r6 & 6 & 2.5   & 0.8 &  250  & fast rotator   \\
r7 & 7 & 2.5   & 0.8 &  250  & fast rotator   \\
r8 & 8 & 2.5   & 0.8 &  250  & fast rotator   \\

\tableline
\end{tabular}
\end{center}
\end{table}

\begin{table}
\begin{center}
\caption{Primary Stellar Parameters at the End of Computations}
\begin{tabular}{cccccccc}
\tableline\tableline
Model & $M$ (\Mo) & $log L$ (\Lo) & $T_{\rm eff}$ & $R$ (\Ro) &
$<v_{\rm rot}>$ (\kms) &  $P_{\rm orbital}$ (d) \\
\tableline
c7 & 0.819 &  4.134 & 2970 & 441 &  0.03 &  11949 \\
c8 & 0.836 &  4.188 & 3148 & 417 &  0.03 &  15164 \\
r7 & 0.841 &  4.212 & 3164 & 424 &  0.44 &  11601 \\
r8 & 0.816 &  4.265 & 3159 & 452 &  0.09 &  15631 \\
\tableline
\end{tabular}
\end{center}
\end{table}

\begin{table}
\begin{center}
\caption{Primary Stellar Parameters when Roche Lobe Overflow is approached}
\begin{tabular}{cccccccc}
\tableline\tableline
Model & $M$ (\Mo) & $log L$ (\Lo) & $T_{\rm eff}$ & $R$ (\Ro) &
$<v_{\rm rot}>$ (\kms) &  d (\Ro) & $P_{\rm orbital}$ (d) \\
\tableline
c5 & 2.206 &  4.147 & 3076 & 415 &  6.1 &  1151.53 & 2612.62 \\
c6 & 1.869 &  4.226 & 2887 & 518 &  4.9 &  1480.16 & 4040.97 \\
r5 & 2.239 &  4.149 & 3086 & 415 &  5.8 &  1144.73 & 2575.88 \\
r6 & 1.999 &  4.239 & 2923 & 513 &  4.8 &  1443.27 & 3799.25 \\
\tableline
\end{tabular}
\end{center}
\end{table}

\clearpage

\begin{figure}
\epsscale{1.10}
\plotone{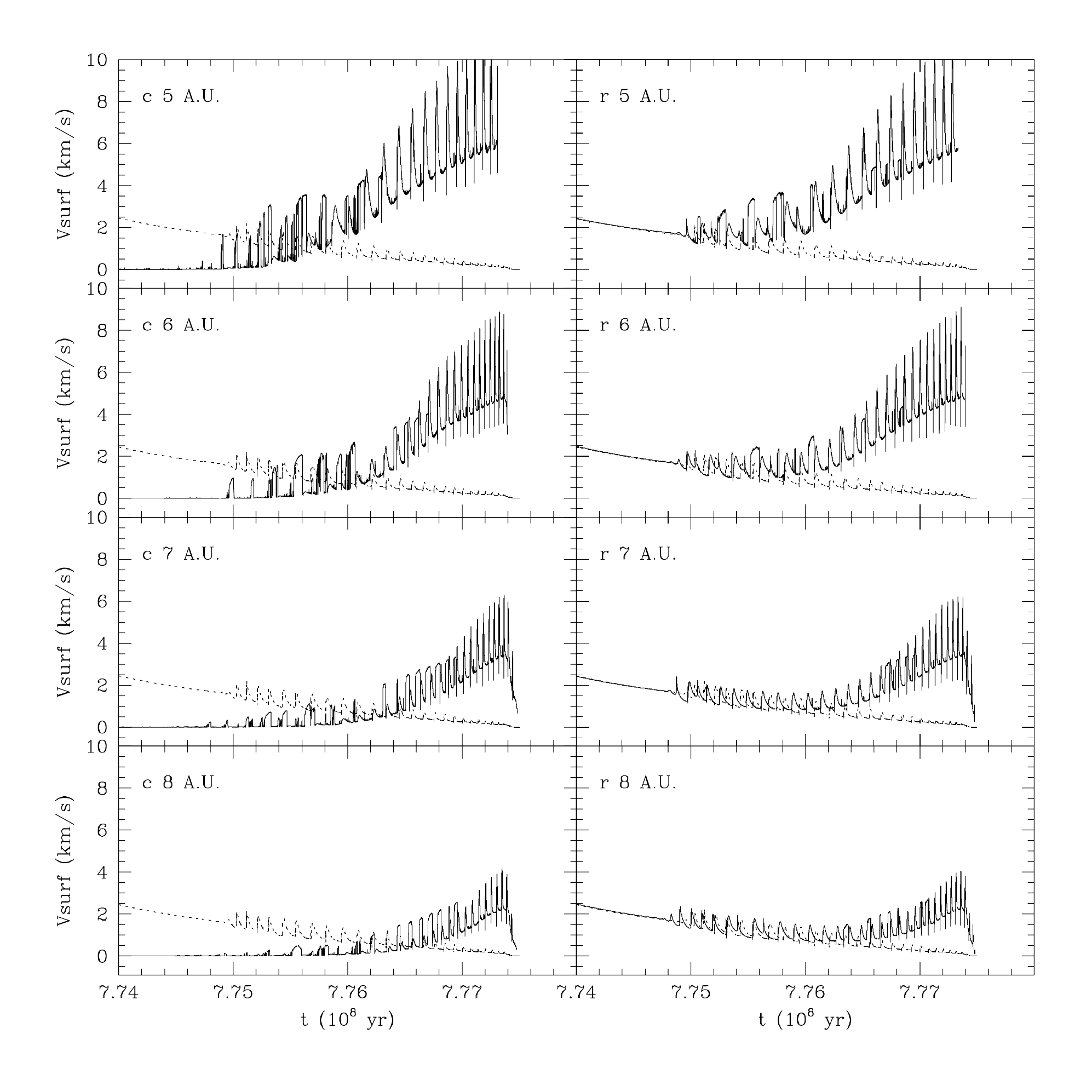}
\caption{Evolution of the surface equatorial rotational velocity
during the thermal pulsing AGB phase.  Left: models with ZAMS velocities nearly zero,
c5, c6, c7 and c8.  Right: models with ZAMS velocities of 250 \kms , r5, r6, r7 and r8.
The initial binary separation is labeled in the upper left corner.
The dotted lines correspond to the magnetic model of paper I as a reference.  }
\label{HR}
\end{figure}

\clearpage

\begin{figure}
\epsscale{1.10}
\plotone{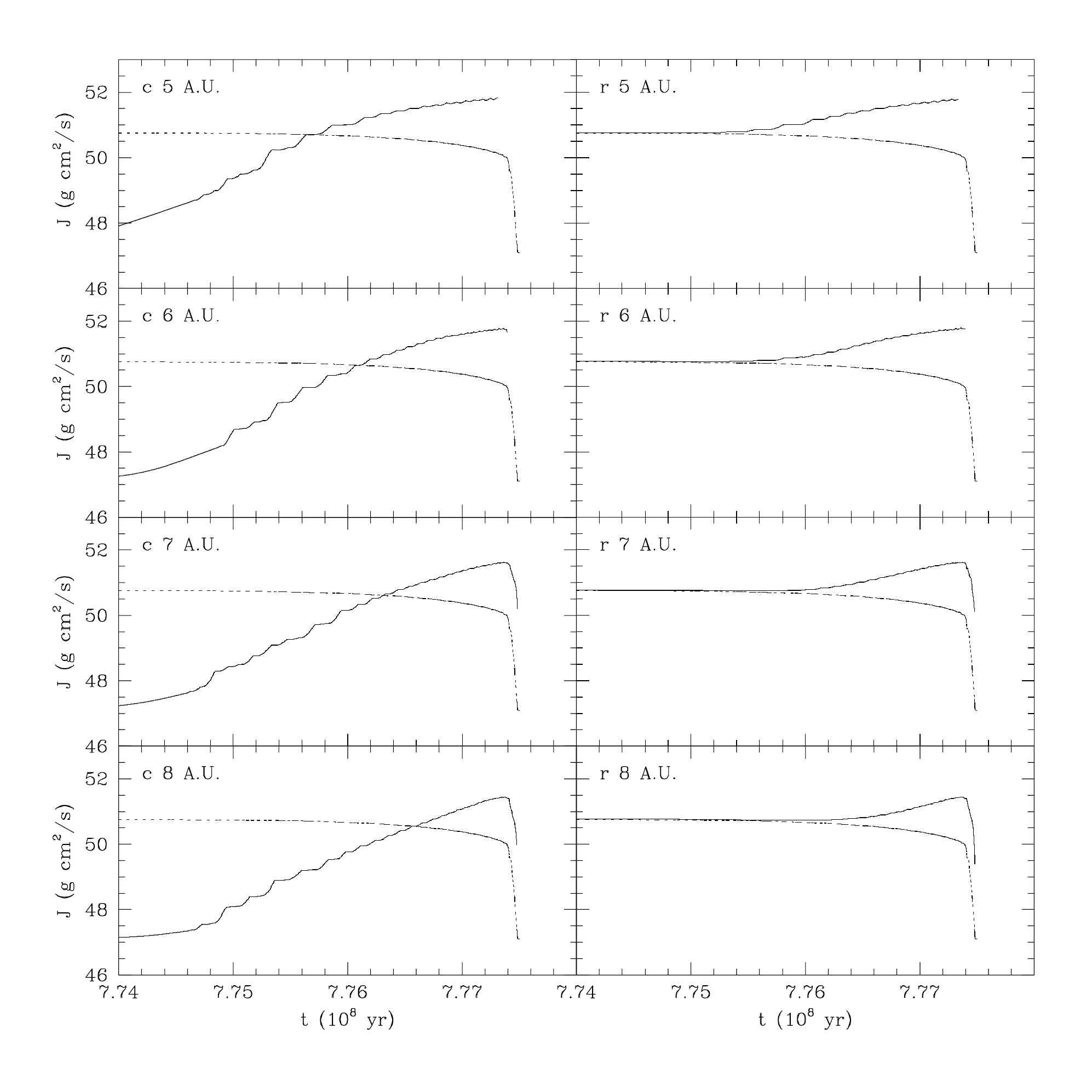}
\caption{Same as figure 1 for the evolution of the spin angular momentum}
\label{HR}
\end{figure}

\clearpage

%

\begin{figure}
\epsscale{1.150}
\plotone{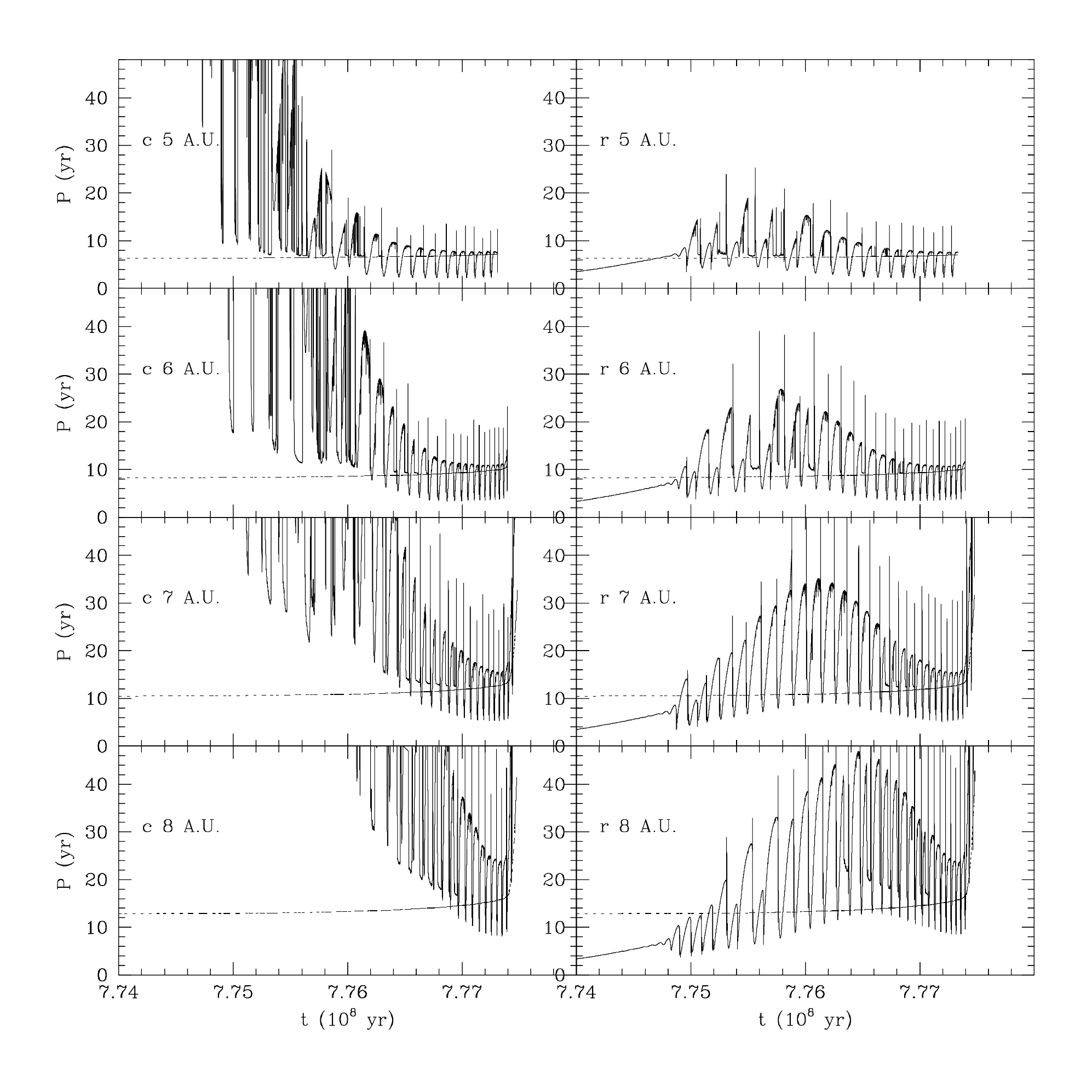}
\caption{Evolution of the spin period (solid line) and orbital period
  (dotted line)
during the thermal pulsing AGB phase for models c5, c6, c7 and c8
(left panels)
and models r5, r6, r7 and r8 (right).}
\label{vsuf}
\end{figure}

\clearpage

\end{document}